\documentclass[]{aa}
\usepackage{amsmath,graphicx,amssymb}
\usepackage{txfonts}
\usepackage{natbib}
\usepackage{hyperref}
\usepackage{siunitx}
\usepackage{textgreek}

\DeclareSIUnit\mag{mag}
\DeclareSIUnit\cycles{c}

\newcommand{\Teff}{$T_\mathrm{eff}$}
\newcommand{\logg}{$\log g$}
\newcommand{\KIC}[1]{KIC\,#1}

\title{Apparent non-variable stars from the Kepler mission}

\author{E.~Paunzen\inst{1,2}
  \and F.~Binder\inst{1}
  \and A.~Cyniburk\inst{1}
  \and M.N.~Duffek\inst{1}
  \and F.~Haberhauer\inst{1}
  \and C.~Heinreichsberger\inst{1}
  \and H.~Kohlhofer\inst{1}
  \and L.~Kueß\inst{1}
  \and H.M.~Maitzen\inst{1}
  \and T.~Saalmann\inst{1}
  \and A.M.~Schanz\inst{1}
  \and S.~Schauer\inst{1}
  \and K.~Schmidt\inst{1}
  \and A.~Tokareva\inst{1}
  \and I.~Wizani\inst{1}}
\institute{Department of Astrophysics, Vienna University, T{\"u}rkenschanzstraße 17, 1180 Vienna, 
Austria
\and Department of Theoretical Physics and Astrophysics, Masaryk University,
Kotl\'a\v{r}sk\'a 2, 611\,37 Brno, Czechia\\
\email{epaunzen@physics.muni.cz}
}

\date{}

\abstract
{The analysis of non-variable stars is generally neglected in the literature. However, such objects are needed for many calibration processes and for testing pulsational models. The photometric time series of the \emph{Kepler} satellite mission still stand as the most accurate data available today and are excellently suited to the search for non-variable stars.}
{We analysed all long-cadence light curves for stars not reported as a variable so far from the  \emph{Kepler} satellite mission. Using the known characteristics and flaws of these data sets, we defined three different frequency ranges where we searched for non-variability.}
{We used the Lomb-Scargle periodogram and the false-alarm probability (FAP) to analyse the cleaned data sets of \num{138451} light curves. We then used $\log FAP \geq -2$ to define a star as 'non-variable' in the ranges below 0.1\,c/d, 0.1 to 2.0\,c/d, and 2.0 to 25.0\,c/d, respectively. Furthermore, we also calculated the standard deviation of the mean light curve to obtain another parameter.}
{In total, we found \num{14154} stars that fulfil the set criteria. These objects are mostly cooler than the \qty{7000}{K} populating the whole main sequence (MS) to the red giant branch (RGB).}
{}
\keywords{Stars: variables: general -- Hertzsprung-Russell and C-M diagrams -- Catalogs -- Methods: data analysis} 

\begin{document}

\maketitle

\titlerunning{Non-variable stars from the Kepler mission}
\authorrunning{Paunzen et al.}

\section{Introduction}

One of the essential questions related to studies aimed at describing stellar objects is whether they are all variable. The most commonly given answer is that all stars are variable, but it is only the amplitude that makes a significant difference. The Sun, the closest star to our planet, is a perfect example \citep{2004A&ARv..12..273F}. Its variability amplitude depends on the wavelength region (with more considerable changes at shorter wavelengths). Furthermore, a timescale of about 11 years, with an amplitude of about \qty{1}{\milli\mag}, a 27-day rotation period (\qty{2}{\milli\mag}), and five-minute short-scale variations (of 0\qty{.15}{\milli\mag}) have been found. An observer from the outside would measure the superposition of all these variations integrated over the solar surface. It also shows how vital time sampling is for detecting variations on different time scales.

In general, variable stars are at the centre  of many scientific studies \citep[e.g.][]{2019A&A...623A.110G}. For example, cepheid variables and their period-luminosity-relations \citep{1912HarCi.173....1L} have allowed us to start constructing a distance ladder, which helped explore large regions of the Universe. The periods, amplitudes, and light curve characteristics \citep{1996lcvs.book.....S} are as manifold as the underlying physical mechanisms \citep{2007uvs..book.....P}.

However, it is important not to lose sight of non-variable stellar objects, as they are very much needed for calibrating absolute fluxes and radial velocities (most photometric variable stars also show strong spectroscopic variations). Techniques such as fitting the spectral energy distribution \citep{2008A&A...492..277B} are crucial depending on the
non-variability. Many photometric calibrations of the effective temperature and metallicity \citep{2006A&A...458..293P, 2017MNRAS.469.3042N} are based on Galactic field stars. However, the variability of the individual stars is generally neglected. For example, the effective temperature of the Cepheid \textkappa{} Pavonis varies between \qtylist{5300; 6300}{\kelvin} over the pulsational phase \citep{2015A&A...576A..64B}. The paper by \citet{2022AJ....163..136M}
investigated the impact of stellar variability in the spectrophotometric
calibration of the science instruments aboard James Webb Space Telescope. They concluded that
variability for cool-type stars cannot be neglected if the full capacities of the instruments
are to be achieved.
A very useful example of how variable stars change their locations in the colour-magnitude diagram is given
in \citet[][in their Figure 11]{2019A&A...623A.110G}. The authors have stated that this figure represents a first step towards
a more global description of stellar variability, namely, by providing new perspectives on the data that can be 
exploited as variable star classification attributes to improve the classification results appreciably. However, it is much more essential to find the astrophysical parameter space that discriminates between variable and non-variable
objects in the same location of the colour-magnitude diagram.
Works analysing non-variable stars are quite rare \citep{1974ApJ...189..293S, 1995BaltA...4..157K, 2001A&A...367..297A}, for two main reasons: 1) establishing non-variability based on a photometric time series is not straightforward and 2) the demand for such objects is not very high. 

We present the results of our analysis of the long-cadence light curves of the \emph{Kepler} mission. We applied a simple cleaning algorithm, starting with the pre-search data conditioning Simple Aperture Photometry (PDCSAP) files. Based on the published experiences with the \emph{Kepler} data sets, we have analysed three different frequency domains and calculated Lomb-Scargle periodograms (LSPs) as well as the false-alarm probabilities (FAPs). These time series data sets are still the most accurate available today.

The final goal is to present a list of non-variable stars in different frequency domains for a given upper limit of the FAP and a mean standard deviation of the cleaned light curve. This set of stars could serve as a basis for calibrating standard fluxes and
the comparison of variable stars in the same region of the HRD.

\section{Overview:\ Kepler Space Telescope}

The light curves used in this work were obtained by the \emph{Kepler Space Telescope}, a successful tool launched in 2009 aimed at discovering extrasolar planets and variable stars, until it was shut down in 2018 due to fuel depletion.

The design of the \emph{Kepler Space Telescope} is pretty simple. It consists of a Schmidt-Telescope with an aperture of \qty{.95}{\meter} and a primary mirror of \qty{1.4}{\meter} in diameter. The light of the stars is collected on a focal plane that consists of 21 pairs of CCDs, each CCD having $2200 \times 1024$ pixels, totalling 95 megapixels on the whole focal plane. The dynamic range of the sensors allowed stars between \qtylist{9; 16}{\mag} to be observed. Those magnitudes were achieved using a broadband white light with wavelengths ranging from \qtyrange{430}{890}{\nano\meter} with a photometric precision of 40 ppm \citep{2010Sci...327..977B}.

Additionally, the telescope was rotated every three months to keep the solar panels in the Sun, allowing the telescope to operate appropriately. These different sets are called `quarters'.
The nine years of the mission were divided into two parts, as described below.

 \emph{Kepler} was the original mission from 2009 to 2013 until two of the four reaction wheels stabilising the telescope failed. A field of view (FOV) of around 115 square degrees in the constellations of Cygnus, Lyra and Draco was chosen for this task. Then, \emph{K2} started in 2014 and observed 20 different FOVs along the ecliptic, each for about 80 days until the telescope ran out of fuel in 2018. \emph{Kepler} used two observing modes for its light curve generation \citep{2012MNRAS.422..665M, 2016ksci.rept....9T}: 1) long cadence (LC): 29.424-minute bins of 270 integrations and 2) short cadence (SC): one-minute bins of 9 integrations.

In total, \emph{Kepler} observed \num{530506} stars during its lifetime. Looking at this huge data set, we have around \num{3000} confirmed exoplanets \citet{2017AAS...22914616A} and about as many published papers.
While the primary goal of the telescope was to find the above-mentioned planets, its precise photometry allowed it to find many new variable stars. It proved very useful for astronomers working in asteroseismology \citet{2010PASP..122..131G} and with binary stars
\citet{2013ApJ...768...33R}. 

\begin{figure}
    \centering
    \includegraphics[width=\columnwidth]{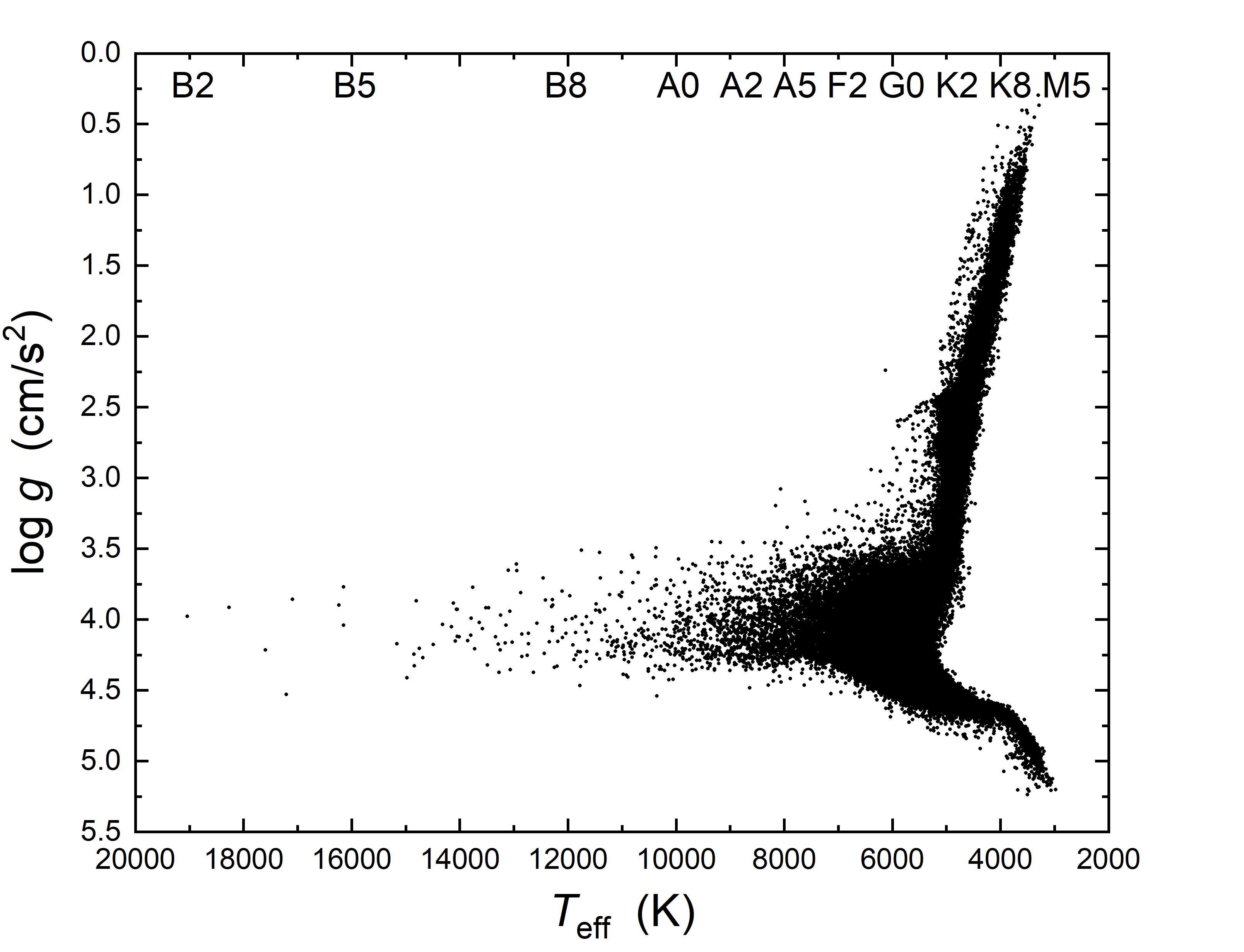}
    \caption{\logg\ versus \Teff\ diagram for the full sample of investigated stars. The astrophysical parameters were taken from \citet{2020AJ....159..280B}.}
    \label{fig:HRD}
\end{figure}

\begin{figure}
    \centering
    \includegraphics[width=\columnwidth]{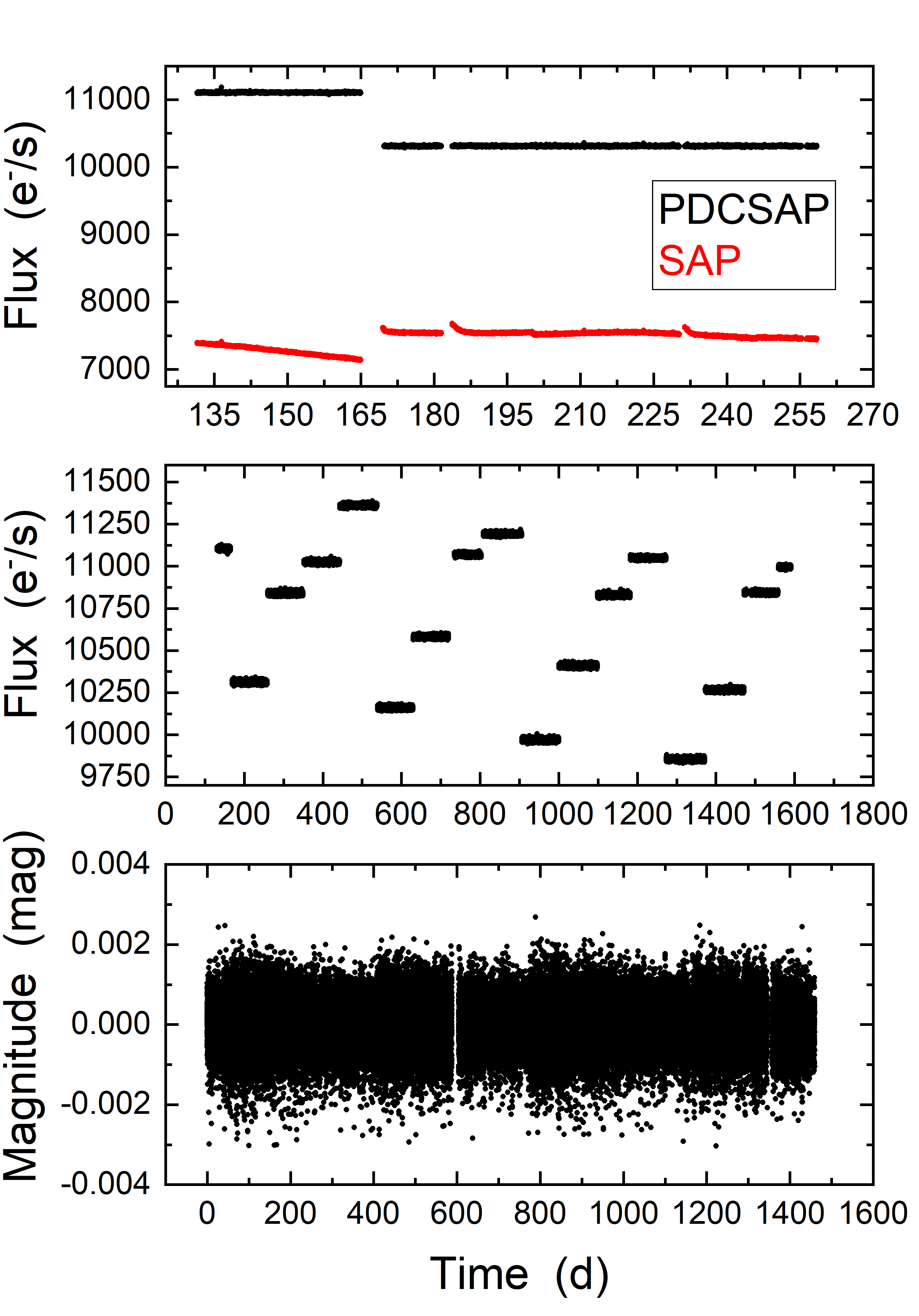}
    \caption{Light curve of the star \KIC{1293468} ($V=\qty{15.545}{\mag}$) observed by \emph{Kepler}. The red and black dots show the PDCSAP and SAP fluxes, respectively. Note: The upper panel only shows two-quarters of the full data set, while the middle panel shows the offsets of the different quarters and the lower panel shows the final cleaned light curve in magnitudes.}
    \label{fig:Example}
\end{figure}

\section{Target selection} \label{target_selection}

We first selected all stars with long cadence light curves available within the original \emph{Kepler} mission for our analysis. We did not consider the about \num{2000} short cadence light curves because they are only a small fraction of the overall data set. 

As the next step, we removed all stars already known as variables. For this, we used the catalogue of the Asteroid Terrestrial-impact Last Alert System \citep[ATLAS,][]{2018AJ....156..241H} and the International Variable Star Index \citep[VSX,][]{2006SASS...25...47W}, along with several works presenting automatic variable star detection routines using \emph{Kepler} data \citep{2010ApJ...713L.204B, 2012AJ....143..101M, 2013ApJ...765L..41S, 2016MNRAS.459.3721B, 2016MNRAS.463.1297Y}. No further constraints were set.
The final sample further processed consisted of \num{138451} light curves. As a next step, we investigated the location of the target stars in the Hertzsprung-Russell diagram (HRD).

\citet{2020AJ....159..280B} presented astrophysical parameters of \num{186301} Kepler stars homogeneously derived from isochrones and broadband photometry, \emph{Gaia} Data Release 2 parallaxes, and spectroscopic metallicities, were available.
However, we wanted to check these values independently. Therefore, we cross-matched our targets with various catalogues to get the distances needed to calculate the bolometric absolute magnitude ($M_\mathrm{Bol}$). The distances were taken from \cite{2021yCat.1352....0B}, where we used the photogeometric values since they seem more reliable. Those are available for \num{106999} of our targets. To calculate the bolometric correction $BC$, we used the polynomials originally published by \cite{1996ApJ...469..355F} and later refined by \cite{2010AJ....140.1158T}. The bolometric corrections were then applied to the absolute \textit{Gaia G} magnitudes ($M_\mathrm{G}$) and the luminosities calculated using $M_\mathrm{Bol,\odot}=\qty{4.75}{\mag}$. We relied on the map of \citet{green2019} for the reddening. The photometric colours \textit{Gaia} $(BP-RP)$ for checking the effective temperatures listed in \citet{2020AJ....159..280B} were also taken from \cite{2021A&A...649A...1G}.

A comparison of our calibrated values and the ones from \citet{2020AJ....159..280B} resulted in an excellent agreement ($\Delta T_\mathrm{eff}$\,=\,3.8\% and
$\Delta M_\mathrm{Bol}$\,=\,2.9\%). 
Such a check is important for the location of stars 
in the HRD, independent of the calibration procedure used.

As shown in Fig.\ref{fig:HRD}, our target star sample consists mainly of objects cooler than \qty{10000}{\kelvin} with most stars being solar type objects, also including a significant amount of objects on the subgiant and red-giant branch. No evolved intermediate and high-mass stars have been included.

\section{Data preparation} \label{data_preparation}

The light curves come in two versions. One is the raw uncorrected flux coming from the telescope (Simple Aperture Photometry, \texttt{SAP}), and the other one has been corrected and already detrended (presearch data conditioning \texttt{SAP}, \texttt{PDCSAP}). For our analysis, we started with the \texttt{PDCSAP} fluxes using only data with optimal quality flags (\texttt{SAP\_QUALITY == 0}). We know further processed light curves exist, such as those from \citet{2011MNRAS.414L...6G} for asteroseismic analyses. However, all those sets are somehow optimised, for instance, certain frequency domains, and we wanted to avoid bias.
The \texttt{PDCSAP} corrects errors, including discontinuities, systematic trends, and outliers,
which obscure the astrophysical signals in the light curves.
It utilises an overcomplete discrete wavelet transform, dividing each light curve into 
multiple channels or bands, thereby allowing for a good separation of characteristic signals and
systematics \citep{2020ksci.rept....8S}. 
The light curves in each band are then corrected utilising a Bayesian maximum 
posterior approach. Some robust fit parameters are then used to generate a Bayesian prior and
a Bayesian posterior probability distribution function (PDF), which (when maximised) finds the
best fit that simultaneously removes systematic effects, while reducing the signal distortion and
noise injection that commonly impacts simple least-squares fittings. A numerical 
and empirical approach was taken where the Bayesian prior PDFs are generated from fits 
to the light-curve distributions themselves \citep{2012PASP..124.1000S}.

The different quarters of one data set have different offsets, which we corrected at first. For this, we calculated the mean and median values for each offset. Then, we compared these values, searching for possible outliers. We used the latter to correct if the mean and median values agree within 2\%. We then applied a simple five-sigma clipping to remove outliers. No further cleaning algorithms were applied to avoid removing any intrinsic variability.

Finally, we transformed the fluxes into magnitudes and subtracted the mean magnitude and the observation starting time. These subtractions guarantee a better numerical performance of the time series algorithm because (quadratic) sums are calculated, which could reach a high 
number for faint magnitudes or large fluxes.
The transformation into magnitudes makes a comparison with the ongoing
ground-based observations such as The All-Sky Automated Survey for Supernovae 
\citep[ASAS-SN;][]{2017PASP..129j4502K}, Optical Gravitational 
Lensing Experiment \citep[OGLE;][]{2015AcA....65....1U}, and 
Zwicky Transient Facility \citep[ZTF;][]{2019PASP..131a8002B} much easier.

Figure \ref{fig:Example} shows the different steps of our data preparation, including the differences between the \texttt{SAP} and \texttt{PDCSAP} fluxes for the faint ($V=\qty{15.545}{\mag}$) star \KIC{1293468} (UCAC4 635-069420). The final cleaned light curve consists of \num{64745} individual data points and has a standard deviation of the mean of \qty{.61}{\milli\mag}.

\begin{figure*}[]
    \begin{center}    
    \includegraphics[width=0.88\textwidth]{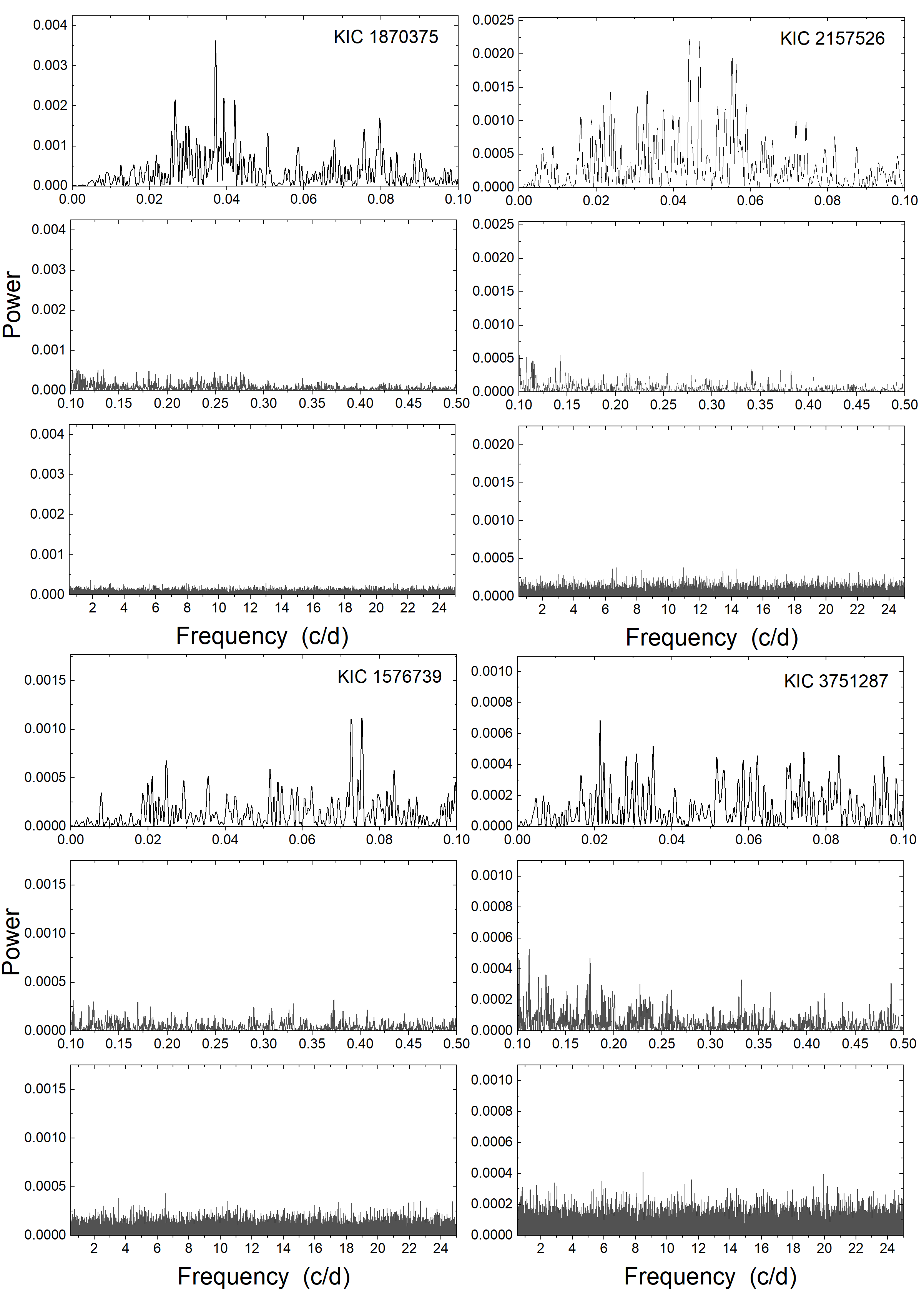}
    \end{center}
   \caption{Examples of different stages of calculated FAPs.
   The LSPs of the stars \KIC{1870375} ($\log FAP$\,=\,$-45$/$-2$/$-1$), 
   \KIC{2157526} ($-21$/$-2$/$0$), \KIC{1576739} ($-8$/$0$/$0$) and \KIC{3751287} ($-2$/$-1$/$0$). Note:  absolute values of the power differ.}
   \label{examples_samples}
\end{figure*}

\section{Time series analysis} \label{tsa}

Detecting a strong periodic signal in a time series is usually straightforward. However, when it comes to small amplitudes or the need to statistically prove that a data set consists only of noise, this can become very difficult \citep{1988nlns.book.....P}.

In astronomy, we normally deal with a discrete, unequally spaced, gapped, and finite data set of an independent (time) and dependent (for example, magnitude) variable, including noise. From a mathematical and statistical point of view, those are the most difficult time series to deal with \citep{1997atsa.book.....S}. The literature shows two different time series methods are commonly used: the Fourier and string-length methods. 

The latter is based on very simple assumptions. First, the data are folded into a series of trial periods. The original data are assigned phases for each, then re-ordered in ascending sequence. The re-ordered data are examined by inspecting them across the full phase interval between zero and one. For each trial period, the sum of the lengths of line segments joining successive points (the string-lengths) are calculated \citep{1965ApJS...11..216L}. In the plot of the string-length versus the trial period, the minima can be considered to correspond to the underlying period. The methods are especially useful for a very small number of randomly spaced observations \citep{1978A&A....63..125R}. 

The Fourier transform connects the time and frequency domains \citep{1976fats.book.....B}. Normally, the power spectrum of a time series, which is the square of the amplitude of each harmonic, is used as a diagnostic tool. This provides each harmonic's contribution to the time series's total energy. 

We used the Lomb-Scargle (LS) algorithm \citep{2009A&A...496..577Z} because it also includes a FAP estimation \citep{2018ApJS..236...16V}. The method is a variation of the discrete Fourier transform (DFT), in which an unequally spaced time series is decomposed into a linear combination of sinusoidal and cosinusoidal functions. The data are transformed from the time to the frequency domain (LSP), which is invariant to time shifts. From a statistical point of view, the resulting periodogram is related to the $\chi^{2}$ for a least-squares fit of a single sinusoid to data, which can treat heteroscedastic measurement uncertainties \citep{1976Ap&SS..39..447L,1982ApJ...263..835S}.

To define whether a time series includes a periodic signal (or not), we followed the statistical approach published by \citet{2008MNRAS.385.1279B}. The FAP measures the probability that a data set with no signal would lead to a peak of a similar magnitude. Previously \citet{1982ApJ...263..835S} had already estimated the cumulative probability of observing a periodogram value in data consisting only of Gaussian noise. \citet{2008MNRAS.385.1279B} improved these analytic estimations based on extreme value theory.

The FAP depends on the frequency domain and its intrinsic noise level. It is advantageous to define different frequency ranges and calculate the FAP.

\section{Non-variability} \label{non_variability}

In general, non-variability is challenging to define, as the final `constancy level' of a time series depends on: 1) the frequency range, namely: the time series can behave differently in the low- and
    high-frequency domain; 2) the time basis of the observations: the spectral window function of the Fourier transform depends on the time basis and gaps of the data; 3) amplitude-noise level: the significance of a detected amplitude depends on the noise level of the corresponding frequency domain; 4) the wavelength region-filter: it is known that the amplitude of variability can depend on the wavelength (see more on classical pulsating variables); 5) applied `pipeline software': it is important to know whether the pre-reduction steps include smoothing in certain frequency domains, such as a low-passband filter; 6) applied `cleaning software': removing possible outliers can significantly alter the detection of significant peaks in the amplitude spectrum; and 7) finally, the applied time series analysis method: different methods are specially developed to detect certain features in the light curve, like transients or flares.

In the case of the \emph{Kepler} mission, this would mean that starting from the \texttt{SAP} data sets, depending on which software and algorithm are used, it is possible to arrive at very different results about a star's non-variability. It is therefore very important to clearly describe each consecutive step of the working flow.
In Sections \ref{data_preparation} and \ref{tsa}, we present the data preparation procedure and the applied time series analysis method. Below, we describe the definition of non-variability we applied for the final data sets.

The general characteristics of the \emph{Kepler} data sets are extensively described in \citet{2012MNRAS.422..665M}. Let us recall that the time basis is three and a half years.
For the LC data, there are especially two problematic frequency regions, at very low frequencies and close to the Nyquist frequency (equal to half the sampling frequency of the light curve, i.e. \qty{25}{\cycles\per\day}). The latter is not important for our analysis because it only affects variability at even higher frequencies. Those frequencies are mirrored and misidentified in the domain close to \qty{25}{\cycles\per\day}, respectively.

As \citet{2012MNRAS.422..665M} pointed out, peaks detected in the low-frequency domain (up to \qty{0.5}{\cycles\per\day}) can result from long time-scale processes such as differential velocity aberration, with stars moving across the CCD, or the changing amount of background contamination light. Also, CCD degradation due to high-energy cosmic ray impacts could, for example, cause such effects. Therefore, the frequency spectrum needs to be interpreted with care.

Based on the instrumental characteristics and the known periods of variable stars \citep{2007uvs..book.....P}, we chose three different frequency domains for which we calculated the LSPs and the FAPs for the highest peak. These are Domain 1 -- below \qty{.1}{\cycles\per\day}; Domain 2 -- \qtyrange{.1}{2.0}{\cycles\per\day}; Domain 3 -- \qtyrange{2.0}{25.0}{\cycles\per\day}.

\begin{figure}
    \centering
    \includegraphics[width=\columnwidth]{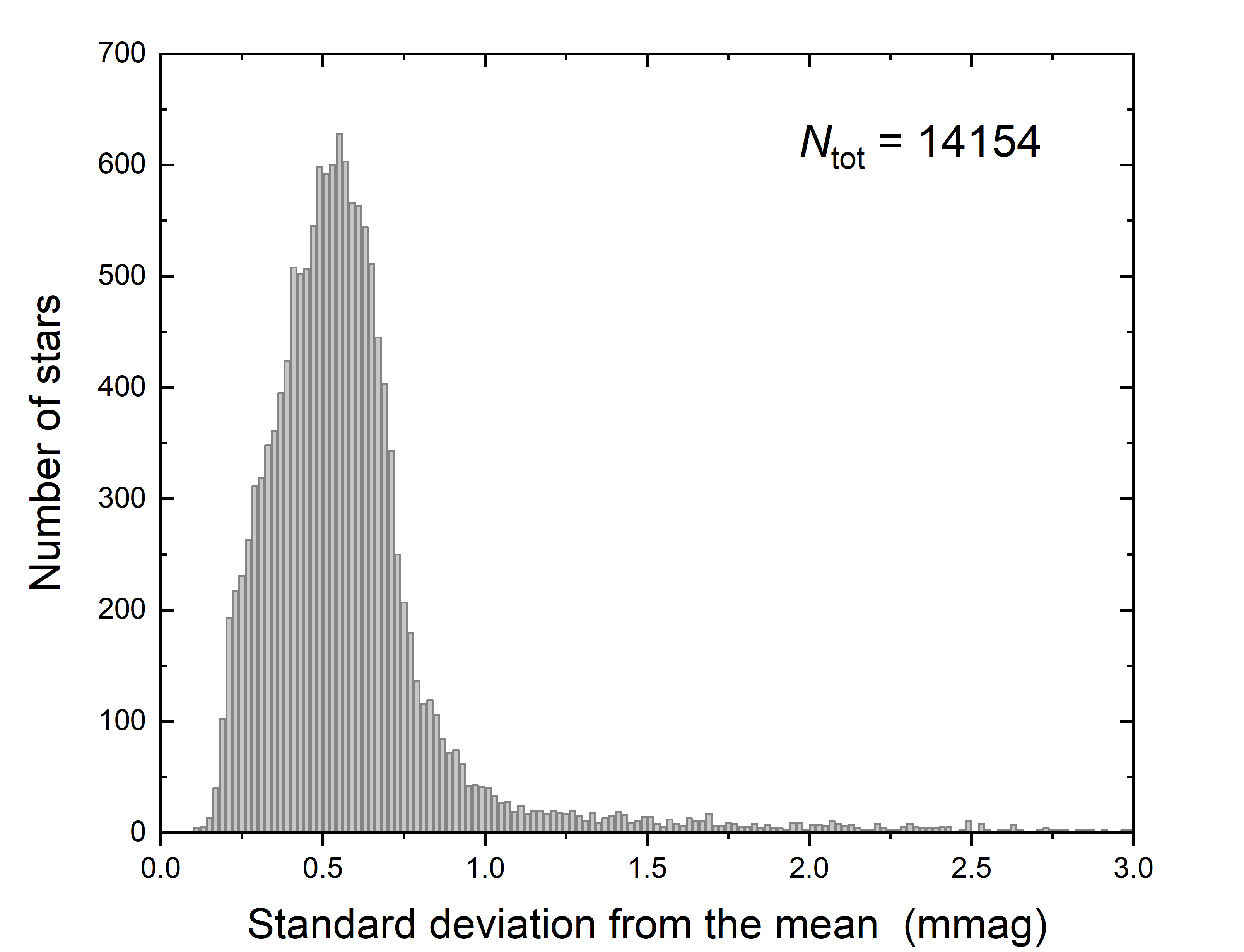}
    \caption{The histogram of the standard deviation of the mean for the combined Sample 1 and 2, respectively.}
    \label{fig:histogram_sigma}
\end{figure}

The first domain includes the most severe instrumentally induced frequencies. We note that the \emph{Kepler} satellite was in an Earth-trailing orbit with a period of 372.5 days. The second and third regions include rotationally induced variability and classical pulsators.

Finally, we had to define when to consider a light curve not included as periodical signal for a given amplitude (or noise level). Using the results from the literature for (in particular) low-amplitude variables of different spectral types \citep{2013MNRAS.430.2313C, 2016MNRAS.460.1970B, 2017MNRAS.472.1618G, 2018A&A...616A..94V, 2018A&A...619A..98H}, we decided to use a value (limit) of $\log FAP \geq -2$ for our subsequent analysis.

In addition to the described time series analysis, we calculated the standard deviation of the mean for each final light curve. This provides a valuable parameter for the data's noise level.
Table \ref{table_master1} lists the essential data for our sample stars (note: full table is only
available in electronic form).

\begin{table*}
\caption{Essential data for our sample stars, sorted by increasing right ascension. The columns denote: 
(1) KIC ID; (2) Right ascension (J2000; $Gaia$ EDR3); (3) Declination (J2000; $Gaia$ EDR3); 
(4) \Teff\ from \citet{2020AJ....159..280B}; (5) \logg\ from \citet{2020AJ....159..280B}; 
(6) Zero point (HJD) of the light curve;
(7) Standard deviation of the mean of the light curve; (8) Logarithmic False-Alarm probability for the frequency range below \qty{.1}{\cycles\per\day};
(9)  Logarithmic False-Alarm probability for the frequency range \qtyrange{.1}{2.0}{\cycles\per\day}; (10) Logarithmic False-Alarm probability for the frequency range \qtyrange{2.0}{25.0}{\cycles\per\day}.}
\label{table_master1}
\begin{center}
\begin{tabular}{cccccccccc}
\hline
\hline
KIC & RA  & DEC  & \Teff  & \logg  & Zero point  & $\sigma$  & $\log FAP_1$  & $\log FAP_2$  & $\log FAP_3$  \\
& (deg) & (deg) & (K) & (cm/s$^{2}$) & (HJD) & (mag) \\
\hline
2141466 &       286.279708      &       37.580639       &       5730    &       4.264   &       2454964.513012  &       0.000759        &       -23     &       -35     &       0       \\
2154980 &       290.232167      &       37.598083       &       5595    &       3.953   &       2454964.512726  &       0.000377        &       -28     &       -9      &       -1      \\
2155278 &       290.309958      &       37.562056       &       5038    &       4.544   &       2455093.245448  &       0.000976        &       -45     &       -12     &       0       \\
2156425 &       290.578667      &       37.564861       &       6446    &       3.901   &       2454964.512703  &       0.000422        &       -20     &       -15     &       -1      \\
2156550 &       290.610750      &       37.521556       &       5812    &       4.406   &       2455185.396621  &       0.000882        &       -39     &       -2      &       0       \\
2156693 &       290.644833      &       37.586694       &       5851    &       4.110   &       2455185.396627  &       0.000825        &       -44     &       -8      &       -2      \\
2156720 &       290.652167      &       37.531333       &       5549    &       4.268   &       2455185.396623  &       0.000877        &       -11     &       -3      &       0       \\
6947741 &       291.532292      &       42.484917       &       6129    &       4.084   &       2455739.866524  &       0.000227        &       -1      &       -4      &       0       \\
6947752 &       291.536208      &       42.459667       &       5287    &       3.669   &       2456206.508701  &       0.000275        &       -5      &       0       &       -2      \\
6947768 &       291.542333      &       42.499861       &       5504    &       3.936   &       2456206.508700  &       0.000234        &       -26     &       0       &       -1      \\
6947932 &       291.596917      &       42.441361       &       5489    &       3.878   &       2454964.512407  &       0.000469        &       -13     &       -3      &       -1      \\
6948287 &       291.723792      &       42.406167       &       5662    &       3.828   &       2454964.512400  &       0.000430        &       -15     &       -3      &       0       \\
6948566 &       291.812292      &       42.440389       &       5720    &       4.316   &       2454964.512393  &       0.000402        &       -12     &       0       &       0       \\
6948885 &       291.904458      &       42.453194       &       5103    &       4.557   &       2454964.512386  &       0.000585        &       -11     &       -1      &       0       \\
6948976 &       291.931250      &       42.428444       &       5370    &       4.083   &       2454964.512385  &       0.000437        &       -3      &       0       &       0       \\
6949090 &       291.972833      &       42.492056       &       6122    &       4.286   &       2456107.160275  &       0.000295        &       -11     &       -1      &       -1      \\
6949102 &       291.975583      &       42.415000       &       5873    &       4.304   &       2454964.512383  &       0.000282        &       -22     &       -2      &       0       \\
6949415 &       292.073042      &       42.474583       &       5914    &       4.437   &       2454964.512374  &       0.000446        &       -18     &       -5      &       0       \\
6949431 &       292.079292      &       42.447667       &       5546    &       4.144   &       2454964.512375  &       0.000521        &       -39     &       -6      &       0       \\
6949539 &       292.114333      &       42.498167       &       5648    &       4.387   &       2454964.512370  &       0.000335        &       -18     &       -2      &       -2      \\
6949702 &       292.163167      &       42.435306       &       6007    &       3.562   &       2455002.765460  &       0.000289        &       -8      &       0       &       0       \\
6949768 &       292.185792      &       42.455583       &       5533    &       4.074   &       2454964.512367  &       0.000540        &       -33     &       -3      &       0       \\
6949871 &       292.217708      &       42.429972       &       6208    &       4.283   &       2454964.512366  &       0.000360        &       -48     &       -23     &       -2      \\
6949873 &       292.218083      &       42.464528       &       6298    &       4.402   &       2455002.765455  &       0.000545        &       -34     &       -5      &       -1      \\
6949891 &       292.225708      &       42.488528       &       5776    &       4.413   &       2454964.512363  &       0.000660        &       -14     &       0       &       0       \\
\hline
\end{tabular}    
\end{center}
\end{table*}

\begin{figure}
    \centering
    \includegraphics[width=\columnwidth]{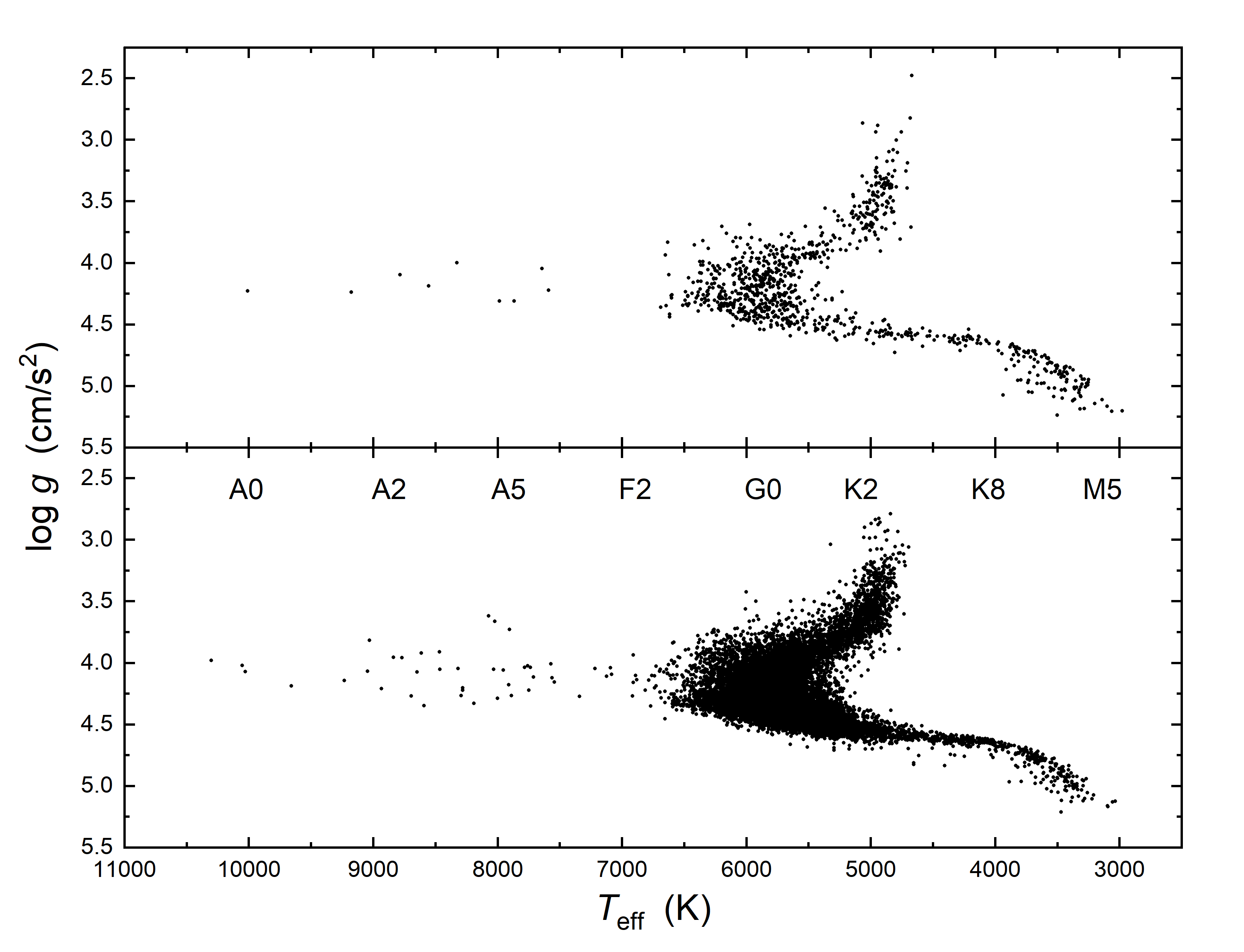}
    \caption{\logg\ versus \Teff\ diagram for sample 1 (upper panel) and sample 2 (lower panel). The astrophysical parameters were taken from \citet{2020AJ....159..280B}.}
    \label{fig:HRD_non_variables}
\end{figure}

\section{Results and conclusions}

As described in Section \ref{target_selection}, we performed a cleaning process and time series analysis of the light curves for \num{138451} stars. According to the limits for non-variability (listed in Section \ref{non_variability}), we defined two samples:

\begin{itemize}
    \item Sample 1: Due to the discussed possible instrumental effects in the low frequency 
    domain, this sample consists of \num{13211} stars with $\log FAP \geq -2$ in the 
    other two domains (\qtyrange{.1}{25.0}{\cycles\per\day}).
    \item Sample 2: This stricter set comprises of \num{943} stars for which $\log FAP \geq -2$ is true for all three frequency domains.
\end{itemize}

In Fig. \ref{examples_samples}, we show four examples of LSPs with different $\log FAP$ values. It clearly shows how difficult it is to interpret the FAPs in the low frequency domain. It is therefore essential to also take the standard deviation of the whole data set and the noise level in the corresponding frequency domain into account. 

If we analyse the histogram of the standard deviation of the mean (Fig. \ref{fig:histogram_sigma}), we find the peak of the (assumed Gaussian) distribution at \qty{0.520(2)}{\milli\mag} and a full width at half maximum (FWHM) of \qty{0.411(4)}{\milli\mag}. Here, we see the effects of the photon noise and the actual observed time basis (number of quarters), namely, the time sampling. It is not immediately possible to transform the noise of the light curve into the noise in the corresponding LSP \citep{2018ApJS..236...16V}. Therefore, the user is advised to inspect the data sets manually.

We recall that we searched for a periodic signal in the data sets. Therefore, the applied 
methods can mainly detect classical pulsation, eclipses, and so on. Single events
such as flares and transits might still be found for stars in our sample.

The catalogue with all essential information will be provided in electronic form in VizieR (CDS). We also set up the project NONVstAR webpage\footnote{\url{https://kepler.physics.muni.cz}}. It includes all light curves and the calculated LSPs. All the corresponding data can be downloaded from there. 

Finally, we investigated the location of the stars of both final samples in the HRD as shown in Figure \ref{fig:HRD_non_variables}. Only a handful of stars are hotter than \qty{7000}{\kelvin}. Otherwise, all evolutionary stages up to the red giant branch (RGB) are well populated. This region exhibits several very different variability classes \citep{2008JPhCS.118a2010E}.

The presented light curves of apparent non-variable stars from the \emph{Kepler} satellite mission are still the most accurate found to date. The applied workflow could easily be adapted to any other datasets, such as those from the Transiting Exoplanet Survey Satellite \citep[TESS,][]{2019MNRAS.490.4040A} mission. As more and more photometric light curves over longer timespans become available, it will be interesting to investigate which astrophysical parameters are distinguishable among variable and non-variable stars in the same location on the HRD. 

\begin{acknowledgements}
    We would like to thank the referee, Susan Mullally, for carefully reading our manuscript and giving such constructive comments, which substantially helped improve the quality of the paper. This paper includes data collected by the Kepler mission and obtained from the MAST data archive at the Space Telescope Science Institute (STScI). Funding for the Kepler mission is provided by the NASA Science Mission Directorate. STScI is operated by the Association of Universities for Research in Astronomy, Inc., under NASA contract NAS 5–26555.
     This work has made use of data from the European Space Agency (ESA) mission {\it Gaia} (\url{https://www.cosmos.esa.int/gaia}), processed by the {\it Gaia} Data Processing and Analysis Consortium (DPAC, \url{https://www.cosmos.esa.int/web/gaia/dpac/consortium}). Funding for the DPAC has been provided by national institutions, in particular the institutions participating in the {\it Gaia} Multilateral Agreement.
\end{acknowledgements}

\bibliographystyle{aa}
\bibliography{Kepler_Part_I}

\end{document}